\newcommand{\bra}[1]{\langle #1 \lvert}
\newcommand{\ket}[1]{\lvert #1\rangle}

\newcommand{\Tr}{\operatorname{Tr}}
\newcommand{\Jx}{J_{ex}}
\newcommand{\Hh}{\mathcal{H}}

\documentclass[aps,twocolumn,raggedbottom,prb,showpacs,nobalancelastpage,amssymb,groupedaddress]{revtex4}
\usepackage{graphicx}
\usepackage{amsmath}
\usepackage{amssymb}
\include{nem}
\begin{document}
\title{Hyperfine induced spin and entanglement dynamics in Double 
Quantum Dots: A homogeneous coupling approach}
\author{B.\, Erbe and J.\, Schliemann}
\affiliation{Institut f\"{u}r Theoretische Physik, Universit\"at
Regensburg, 93053 Regensburg, Germany}
\date{\today}

\begin{abstract}
We investigate hyperfine induced electron spin and entanglement dynamics in a 
system of two quantum dot spin qubits. We focus on the situation of
zero external magnetic field and concentrate on
approximation-free theoretical methods.
We give an exact solution of the model for homogeneous hyperfine coupling 
constants (with all coupling coefficients being equal) and varying exchange 
coupling, and we derive the dynamics therefrom. 
After describing and explaining 
the basic dynamical properties, the decoherence time is calculated from the 
results of a detailed investigation of the short time electron spin dynamics. 
The result turns out to be in good agreement with experimental data.
\end{abstract}
 \maketitle

\section{Introduction}
Quantum dot spin qubits are among the most promising and most intensively
investigated building blocks of possible future solid state quantum
computation systems \cite{LossDi98,Hanson07}. 
One of the major limitations of the decoherence time of the confined 
electron spin is its interaction with surrounding nuclear spins by means 
of hyperfine interaction 
\cite{KhaLossGla02,KhaLossGla03,expMarcus,Koppens05,Petta05,Koppens06,Koppens08,Braun05}. For 
reviews the reader is referred to 
Refs.~\cite{SKhaLoss03,Zhang07,Klauser07,Coish09,Taylor07}.
Apart from this adverse aspect, 
hyperfine interaction can act as a resource of quantum information 
processing 
\cite{Taylor03,SchCiGi08,SchCiGi09,ChriCiGi09, ChriCiGi07,ChriCiGi08}. 
For the above reasons it is of key interest to understand the hyperfine 
induced spin dynamics. 

Most of the work into this direction, for single as well as double quantum dots, has been carried out under the assumption of a strong magnetic field coupled to the central spin system. This allows for a perturbative treatment or a complete neglect
 of the electron-nuclear ``flip-flop'' part of the Hamiltonian, yielding great simplification \cite{KhaLossGla02, KhaLossGla03, Coish04, Coish05, Coish06, Coish08}. In the present paper we consider the case of zero magnetic field where such approximations fail, and we therefore concentrate on exact methods.

In the case of a single quantum dot spin qubit the usual Hamiltonian
describing hyperfine interaction with surrounding nuclei is integrable
by means of Bethe ansatz as devised by Gaudin several decades ago\cite{Gaudin,John09,BorSt071,BorSt09}. In the following we shall refer to that sytem also as the Gaudin model.
Nevertheless exact results are rare also here because the Bethe ansatz equations 
are very hard to handle. Hence there are mainly three different 
routes in order to gain some exact results:
(i) Restriction of the initial state to the one magnon sector 
\cite{KhaLossGla02, KhaLossGla03}, (ii)  restriction to small system
sizes enabling progress via exact numerical
diagonalizations \cite{SKhaLoss02,SKhaLoss03},
and (iii) restrictions to the hyperfine coupling constants
\cite{BorSt07, ErbS09}.
In the present paper we will follow the third route and
study in detail the electron spin as well as the entanglement 
dynamics in a double quantum dot model with partially homogeneous 
couplings: The hyperfine coupling constants are chosen to be equal to each other, whereas the exchange coupling is arbitrary.
Although the assumption of homogeneous hyperfine
constants (being the same for each spin in the nuclear bath) is certainly
a great simplification of the true physical situation, models
of this type offer the opportunity to obtain exact, approximation-free
results which are scarce otherwise. Moreover, such models 
have been the basis of several recent theoretical studies
leading to concrete predictions
\cite{SchCiGi08,SchCiGi09,ChriCiGi09,ChriCiGi08}. 

The paper is organized as follows: In Sec. \ref{model} we introduce the Hamiltonian
of the hyperfine interaction and derive the spin and entanglement dynamics for homogeneous 
hyperfine coupling constants. In Sec. \ref{dynamics} we study the spin and entanglement dynamics for different 
exchange couplings and bath polarizations. For the completely homogeneous case of the exchange coupling being the same as the hyperfine
couplings we find an empirical rule describing the transition from low polarization dynamics to high polarization dynamics.
The latter shows a jump in the amplitude when varying the exchange coupling away from complete homogenity. 
This effect as well as features like the periodicity of the dynamics are explained by analyzing the level spacings and their 
contributions to the dynamics. In Sec. \ref{decoherence} we extract the decoherence time from the dynamics by 
investigating the scaling behaviour of the short time electron spin dynamics. The result turns out to be in good agreement with
experimental findings.

\section{Model and formalism}
\label{model}
The hyperfine interaction in a system of two quantum dot spin qubits is described by the Hamiltonian
\begin{equation}
\label{1}
 H= \vec{S}_1 \cdot \sum_{i=1}^N A_i^1 \vec{I}_i 
+ \vec{S}_2 \cdot \sum_{i=1}^N A_i^2 \vec{I}_i + \Jx \vec{S}_1 \cdot \vec{S}_2 ,
\end{equation}
where $\Jx$ denotes the exchange coupling between the two electron spins $\vec S_1$, $\vec S_2$, and 
$A_i^1 $, $A_i^2 $ are the coupling parameters for their hyperfine interaction with the surrounding nuclear spins $\vec I_i$. 
\begin{figure}[h!]
\begin{flushright}
\resizebox{\linewidth}{!}{
\includegraphics{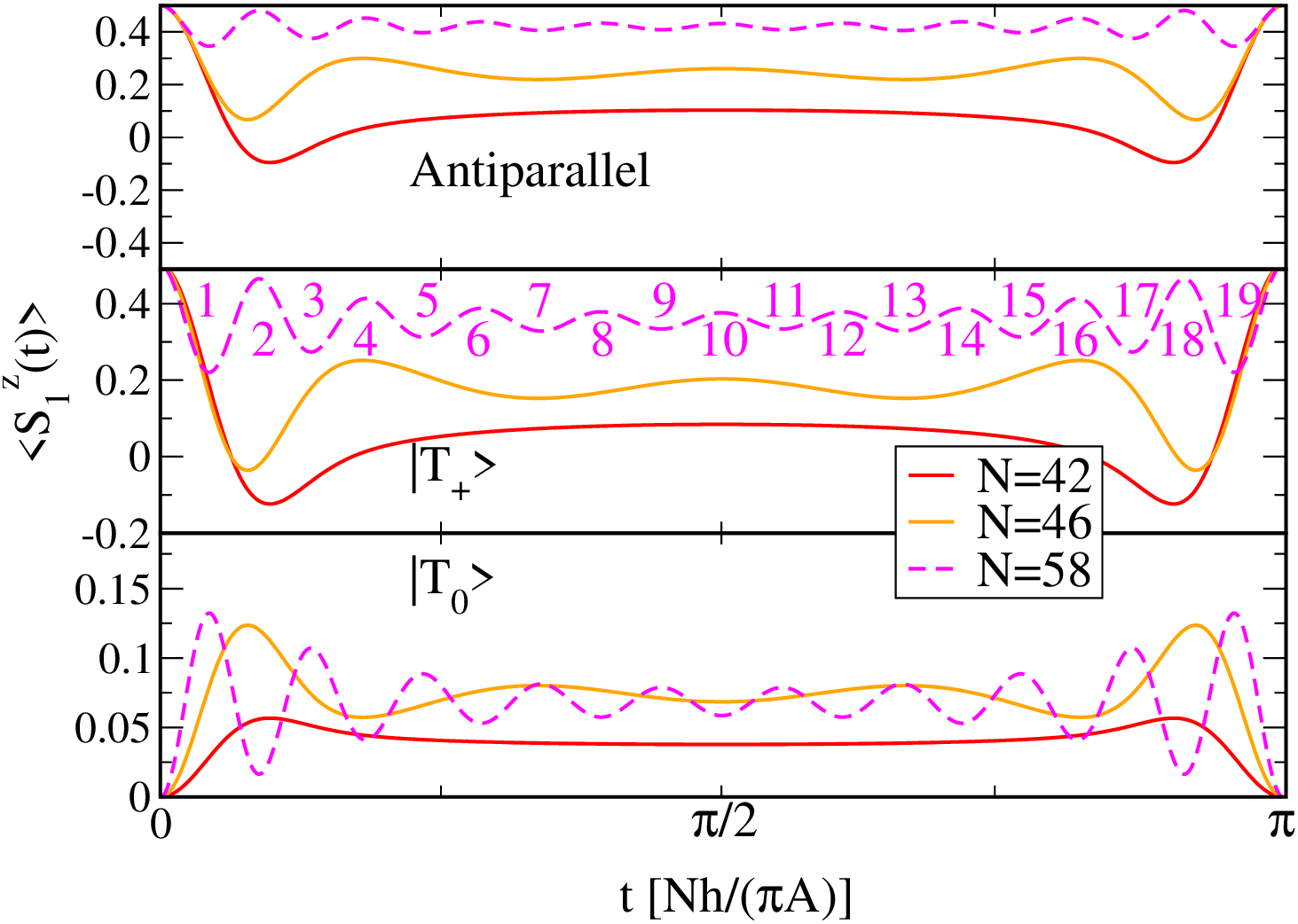}}
\end{flushright}
\caption{\label{Fig:evenodd1} (Color online) Spin dynamics for $\ket{\alpha_1}=\ket{\Uparrow \Downarrow}, \ket{T_+}, \ket{T_0}$ and an even number of spins. The number of down spins in the bath is $N_D=20$ in all plots, yielding polarizations $p_b\approx 5\%-30\%$. Note that the time unit is rescaled according to the number of bath spins. We see periodicity with $\pi$. For $\ket{\alpha_1}=\ket{T_0}$ and $N=58$ we count the number of local extrema on one period and find $N-2N_D+1=58-40+1=19$ as expected.}
\end{figure}
In a realistic quantum dot these quantities
are proportional to the square modulus of the electronic 
wave function at the sites of the nuclei and therefore 
clearly spatially dependent
\begin{equation}
\label{cpl}
A_i^{j}=A_i v \left|\psi^{j}(\vec{r}_i)\right|^2,
\end{equation}
where $v$ is the volume of the unit cell containing one nuclear spin and $\psi^{j}(\vec{r}_i)$ is the electronic wave function of electron $j=1,2$ at the site of $i$-th nucleus. The quantity $A_i$ denotes the hyperfine coupling strength which depends on the respective nuclear species through the nuclear gyromagnetic ratio \cite{Coish09}. It should be stressed that these can have different lengths. In a GaAs quantum dot for example all Ga and As isotopes carry the same nuclear spin $I_i=3/2$, whereas in an InAs quantum dot the In isotopes carry a nuclear spin of $I_i=9/2$ \cite{SKhaLoss03}. In any case the Hamiltonian obviously conserves
the total spin $\vec{J}=\vec{S} + \vec{I}$, where 
$\vec{S}=\vec{S}_1 + \vec{S}_2$ and $\vec{I}=\sum_{i=1}^N \vec{I}_i$. 

The model to be studied in this paper now results by neglecting the spatial variation of the hyperfine coupling constants and choosing them to be equal to each other $A^1_i=A^2_i=A/N$. Variation of the exchange coupling between 
the two central spins $\Jx$ then gives rise to an inhomogeneity in the system. Hence the two electron spins are interacting with a common nuclear spin bath. Moreover, if small variations
of the coupling constants would be included, degenerate energy levels would
slightly split and give rise to a modified {\em long-time} behavior of the
system. In our quantitative studies to be reported on below, however, we 
focus on the {\em short-time} 
properties where decoherence phenomena take place. Indeed, in section 
\ref{decoherence} we
obtain realistic $T_{2}$ decoherence time scales in an almost
analytical fashion.
\begin{figure}[h!]
\begin{flushright}
\resizebox{\linewidth}{!}{
\includegraphics{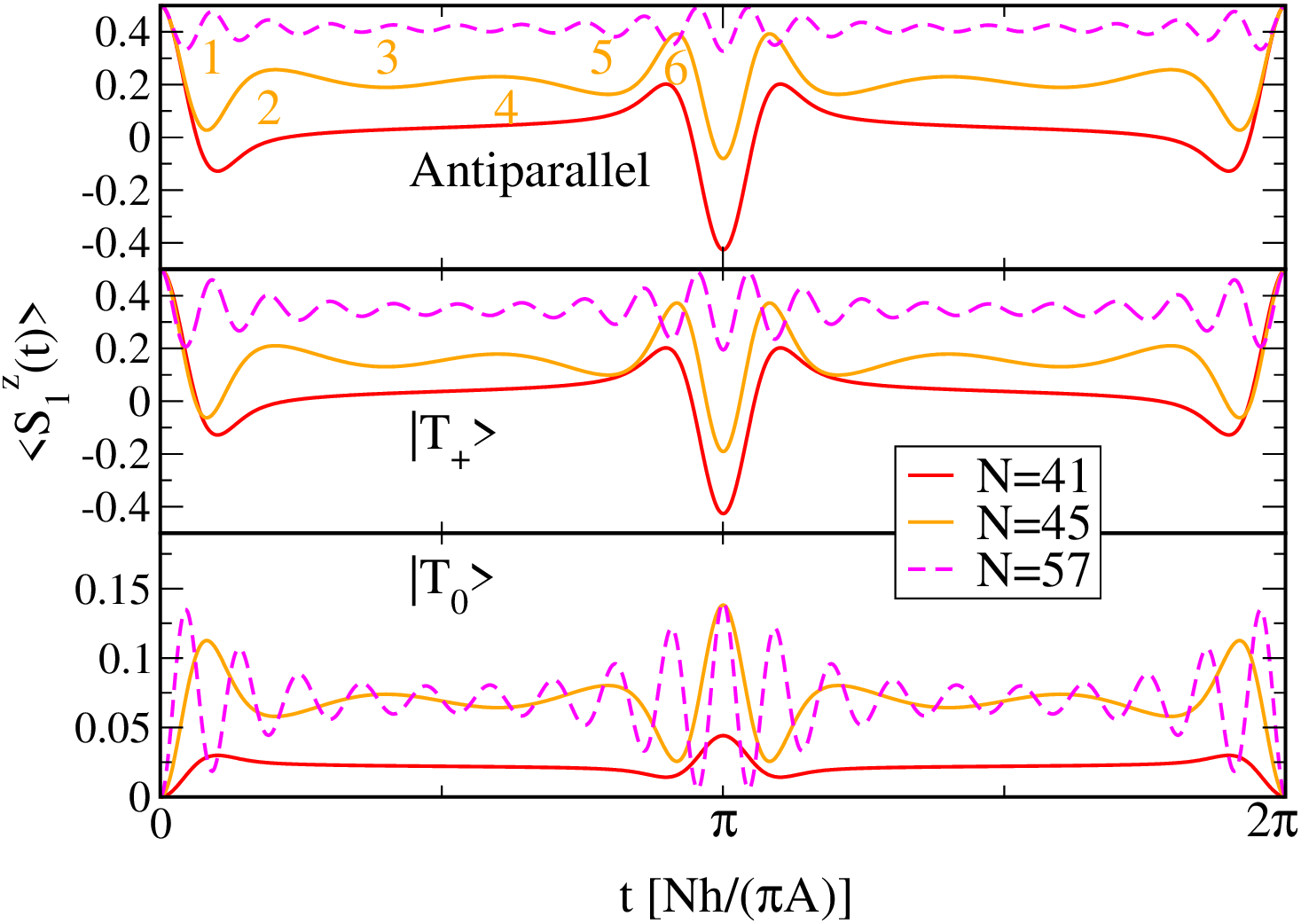}}
\end{flushright}
\caption{\label{Fig:evenodd2} (Color online) Spin dynamics for $\ket{\alpha_1}=\ket{\Uparrow \Downarrow}, \ket{T_+}, \ket{T_0}$ and an odd number of spins. The number of down spins in the bath is $N_D=20$ in all plots, giving polarizations $p_b \approx 2\%-30\%$. In contrast to the case of an even number of spins we see periodicity with $2 \pi$. For $\ket{\alpha_1}=\ket{\Uparrow \Downarrow}$ and $N=45$ we count the number of local extrema on half the period and find $N-2N_D+1=45-40+1=6$ as expected.}
\end{figure}
In consistency with the homogenous couplings we choose the length of the bath spins to be equal to each other. For simplicity we restrict the nuclear spins to $I_i=1/2$. 
We expect our results to be of quite general nature not strongly depeding on this choice \cite{John09}. Note that both, the square $\vec S^{2}$ of the total central spin as well as the square $\vec I^{2}$ of the total bath spin are separately
conserved quantities. 

Considering the two electrons to interact with a common nuclear spin bath 
as in our model corresponds to a physical situation where the electrons 
are comparatively near to each other. This leads to the question whether 
our model 
is also adapted to the case of two electrons in one quantum dot, rather 
than in two nearby quantum dots. Assuming perfect confinement, in the former 
case one of the two electrons would be forced into the first excited state, 
which typically has a zero around the dot center. Thus, the coupling 
constants near the very center of the dot would clearly be different 
for the two electrons. Therefore our model is more suitable for the 
description of two electrons in two nearby quantum dots than for the 
case of two electrons in one dot.

Let us now turn to the exact solution of our homogeneous coupling model and 
calculate the spin and entanglement dynamics from the eigensystem. 
In what follows we shall work in subspaces of a fixed eigenvalue of 
$J^z$. Thus, the expectation values of the 
$x$- and $y$-components of the central and nuclear spins vanish, and we only have to consider their $z$-components. 
\begin{figure}[h!]
\begin{flushright}
\resizebox{\linewidth}{!}{
\includegraphics{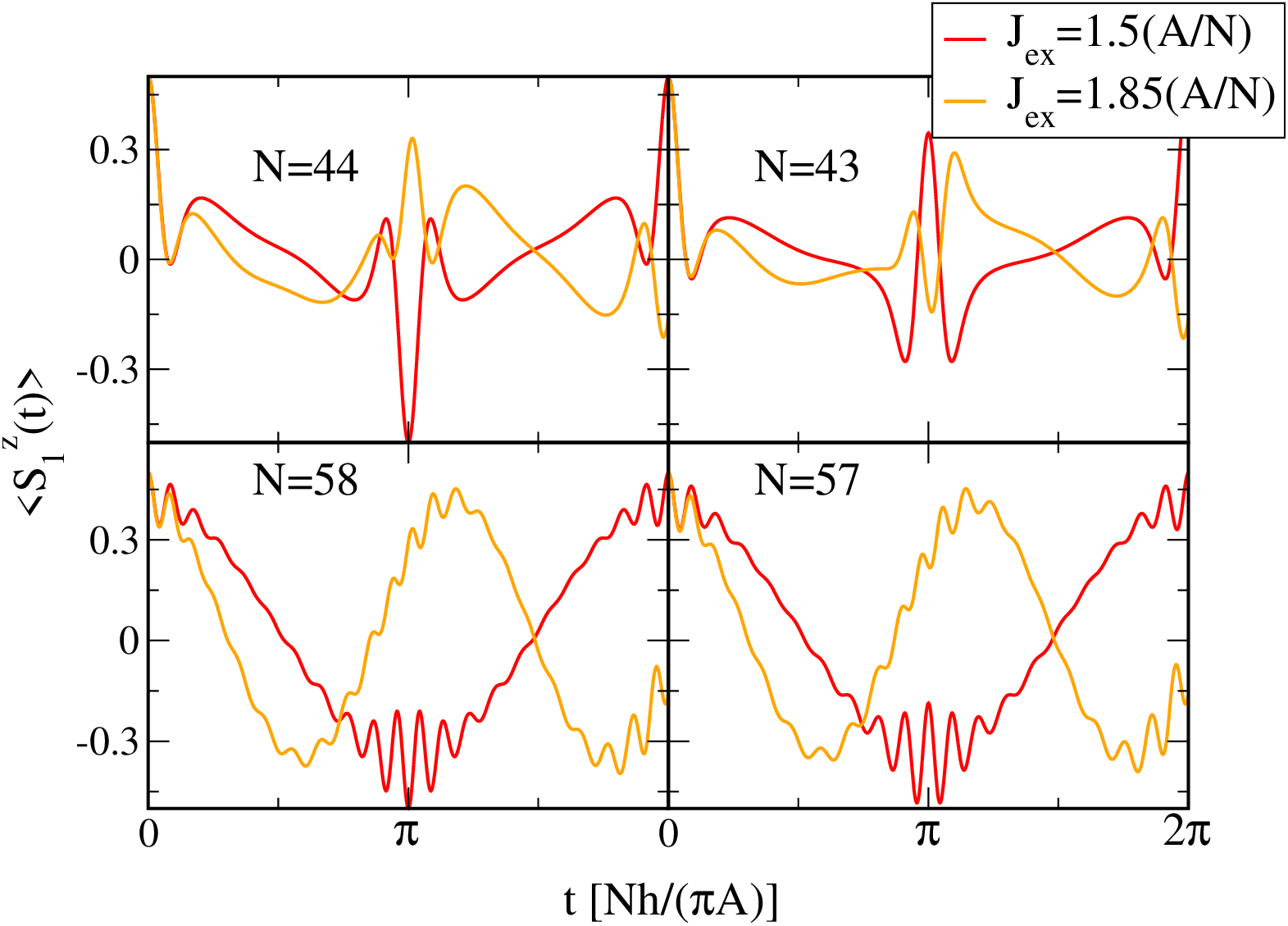}}
\end{flushright}
\caption{\label{Fig:evenoddJ1} (Color online) Spin dynamics for $\ket{\alpha_1}=\ket{\Uparrow \Downarrow}$ and $N_D=20$, resulting in $p_b \approx 6 \% - 30 \%$. If $\Jx$ is an odd multiple of $A/2N$ we see periodicity with $2\pi$. }
\end{figure}

If all hyperfine couplings are equal to each other $A^1_i=A^2_i=A/N$, the 
Hamiltonian (\ref{1}) can be rewritten in the following way
\begin{equation}
\label{5}
H=H_{\operatorname{hom}}+\left(\Jx-\frac{A}{N} \right)\vec{S}_1 \cdot \vec{S}_2
\end{equation}
with
\begin{equation}
\label{2}
H_{\operatorname{hom}}=\frac{A}{2N}\left( \vec{J}^2 - \vec{S}^2_1 
- \vec{S}^2_2 - \vec{I}^2\right).
\end{equation}
Omitting the quantum numbers corresponding to a certain Clebsch-Gordan 
decomposition of the bath, the eigenstates are labelled by $J,m,S$ 
associated with the operators $\vec{J}^2, J^z, \vec{S}^2$. 
The two central spins couple to $S=0,1$. Hence the eigenstates of 
$H$ are given by triplet states $\ket{J,m,1}$, corresponding to the 
coupling of a spin of length one to an arbitrary spin, and a 
singlet state $\ket{J,m,0}$. The explicit expressions are given
 by (\ref{eig1}, \ref{eig2}, \ref{eig3}) in appendix A. 

The corresponding eigenvalues read as follows:
\begin{small}
\begin{subequations}\label{4}
\begin{eqnarray}
H \ket{I+1,m,1}&=&\left( \frac{A}{N}I+\frac{\Jx}{4} \right) 
\ket{I+1,m,1}\\ \label{4b}
H \ket{I,m,1}&=&\left(  \frac{\Jx}{4}-\frac{A}{N}\right) 
\ket{I,m,1}\\ \label{4c}
H \ket{I-1,m,1}&=&\left(-\frac{A}{N}I+\frac{\Jx}{4}-\frac{A}{N} 
\right) \ket{I-1,m,1}\\ 
H \ket{I,m,0}&=& -\frac{3}{4}\Jx  \ket{I,m,0}
\end{eqnarray}
\end{subequations}
\end{small}
Now we are ready to evaluate the time evolution of the central spins and 
their entanglement from the eigensystem of the Hamiltonian. 
We consider initial states $\ket{\alpha}$ of the form 
$\ket{\alpha}=\ket{\alpha_1}\ket{\alpha_2}$, where $\ket{\alpha_1}$ is an 
arbitrary central spin state and $\ket{\alpha_2}$ is a product of $N$ states $\ket{\uparrow},\ket{\downarrow}$. 
\begin{figure}[h!]
\begin{flushright}
\resizebox{\linewidth}{!}{
\includegraphics{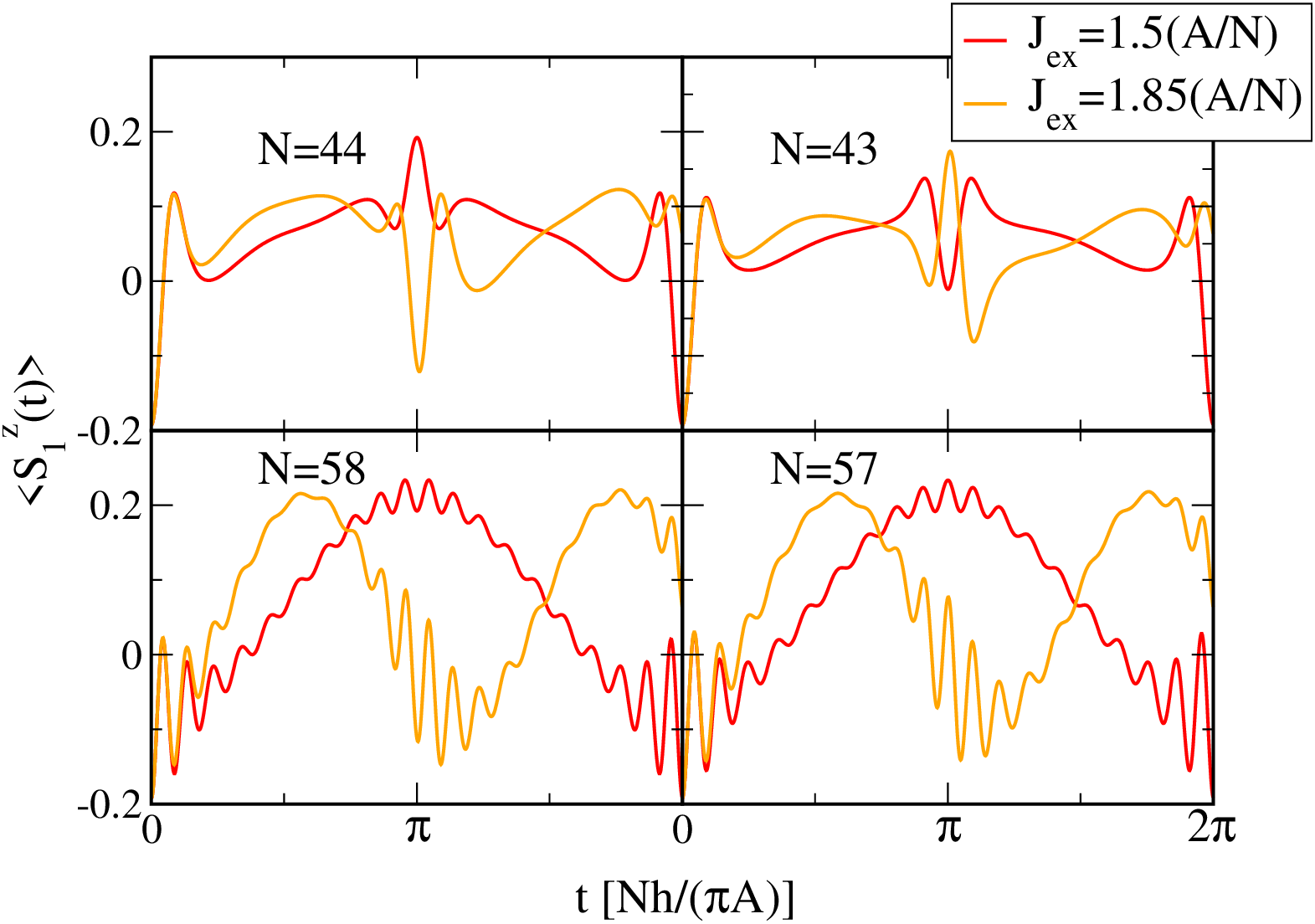}}
\end{flushright}
\caption{\label{Fig:evenoddJ2} (Color online) Spin dynamics for $\ket{\alpha_1}=(1/\sqrt{13})\left(2 \ket{\Uparrow \Downarrow}+3\ket{\Downarrow \Uparrow}\right)$ and $N_D=20$, resulting in $p_b \approx 6 \% - 30 \%$.}
\end{figure}
The physical significance of this choice becomes clear by rewriting the electron-nuclear coupling parts of the Hamiltonian in terms of creation and annihilation operators:
\begin{equation}
\label{flipflop}
\vec{S}_i\vec{I}_j=\frac{1}{2}\left(S_i^+I_j^- + S_i^-I_j^+\right)+S_i^z I_j^z
\end{equation}
Obviously the second term does not contribute to the dynamics for initial states which are simple product states. Hence by considering initial states of the above form, we mainly study the influence of the flip-flop part on the dynamics of the system. This is exactly the part which is eliminated by considering a strong magnetic field like in Refs. \cite{KhaLossGla02, KhaLossGla03, Coish04, Coish05, Coish06, Coish08}. 

As the $2^N$ dimensional bath Hilbert space is spannend by the $\vec{I}^2$ eigenstates, every product state can be written in terms of 
these eigenstates. If $N_D \leq N/2$ is the number of down spins in 
the bath, it follows
\begin{small}
\begin{equation}
\label{8}
\ket{\underbrace{\downarrow \ldots \downarrow}_{N_D} \uparrow \ldots \uparrow}
= \sum_{k=0}^{N_D} \sum_{\left\{S_i\right\}} c_k^{\left\{S_i\right\}} 
\ket{\underbrace{\frac{N}{2}-k}_{I},\frac{N}{2}-N_D,\left\{S_i\right\}},
\end{equation}
\end{small}
where the quantum numbers $\lbrace S_i \rbrace$ are due to a certain 
Clebsch-Gordan decomposition of the bath. In (\ref{8}) we assumed the 
first $N_D$ spins to be flipped, which is no loss of generality due to the 
homogeneity of the couplings. For the following discussions it is convenient to introduce the 
bath polarization $p_b=\left(N-2N_D \right)/N $. 

Using (\ref{8}) and inverting (\ref{eig1}, \ref{eig2}, \ref{eig3}), 
the time evolution can be calculated by writing $\ket{\alpha}$ in 
terms of the above eigenstates and applying the time evolution operator. 
Using (\ref{eig1}, \ref{eig2}, \ref{eig3}) again and tracing out the bath degrees of freedom we arrive at the reduced density matrix $\rho(t)$, which enables to evaluate the expectation value $\langle S^z_{1/2} (t) \rangle$ and the dynamics of the entanglement between the two central spins. 
\begin{figure}[h!]
\begin{flushright}
\resizebox{\linewidth}{!}{
\includegraphics{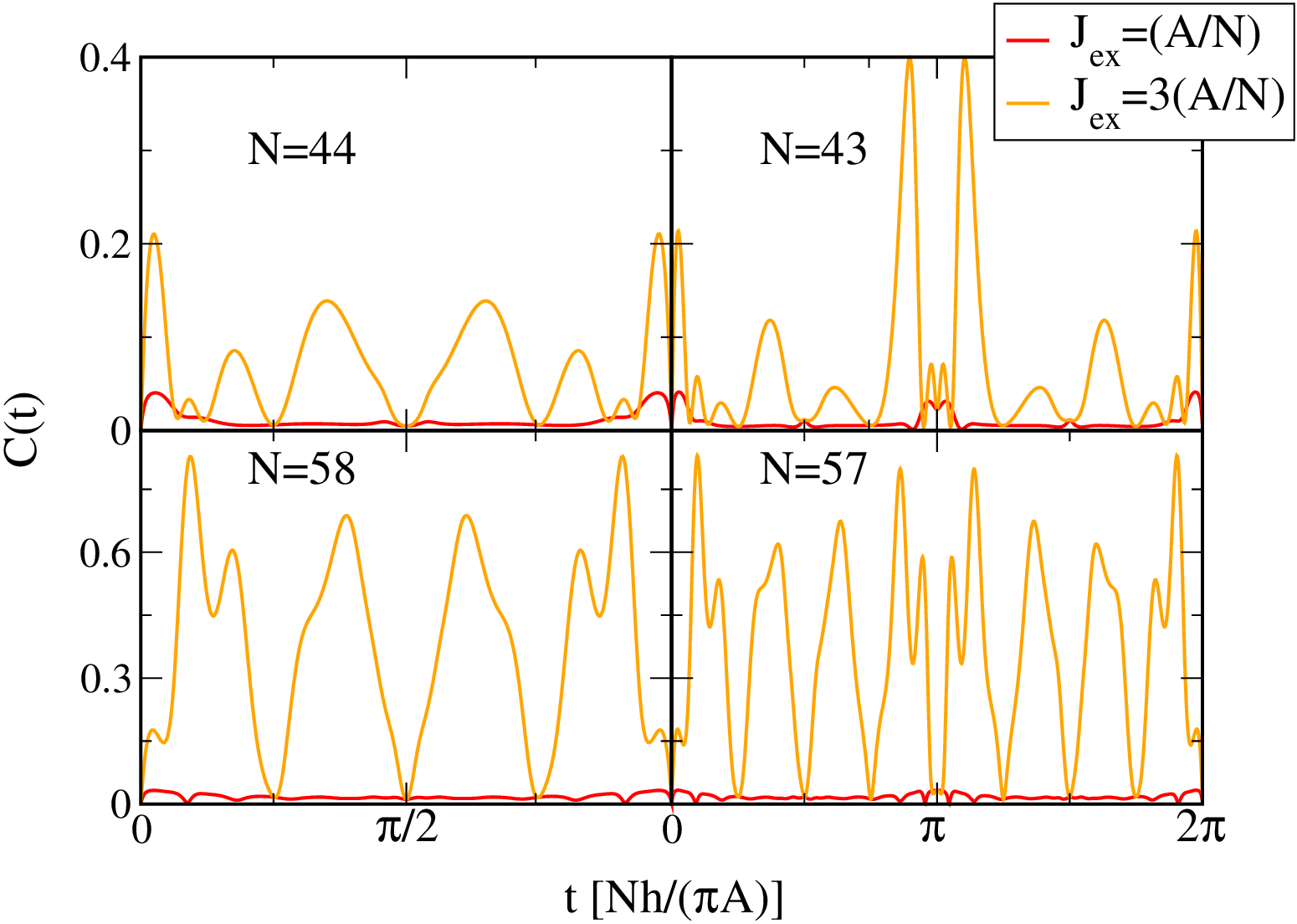}}
\end{flushright}
\caption{\label{Fig:con_hom1} (Color online) Entanglement dynamics for $\ket{\alpha_1}=\ket{\Uparrow \Downarrow}$ and $N_D=20$, resulting in $p_b \approx 6 \% - 30 \%$. In the completely homogeneous case the amplitude is small even for high polarization. Generation of entanglement benefits from high polarization.}
\end{figure}
As a measure for the entanglement we use the concurrence \cite{Wootters97}
\begin{equation}
 C(t)=\operatorname{max}\lbrace0,\sqrt{\lambda_1}-\sqrt{\lambda_2}-\sqrt{\lambda_3}-\sqrt{\lambda_4}\rbrace,
\end{equation}
where $  \lambda_i$ are the eigenvalues of the non-hermitian matrix $\rho(t) \tilde{\rho}(t)$ in decreasing order. Here $\tilde{\rho}(t)$ is given by $\left(\sigma_y \otimes \sigma_y \right)\rho^*(t) \left( \sigma_y \otimes \sigma_y \right) $, where $\rho^*(t)$ denotes the complex conjugate of $\rho(t)$. 
The coefficients $c_k^{\left\{S_i\right\}}$ are of course products of Clebsch-Gordan coefficients, which enter the time evolution through the quantity 
\begin{equation}
d_k=\sum_{\lbrace S_i \rbrace}\left(  c_k^{\lbrace S_i \rbrace}\right)^2
\end{equation}
and usually have to be calculated numerically. The main advantage in considering $I_i=1/2$ is now that in this case a closed expression for $d_k$ can be derived \cite{BorSt07}:
\begin{equation}
\label{10}
d_k =\frac{N_D!(N-N_D)!(N-2k+1)}{(N-k+1)!k!}
\end{equation} 

For further details on the calculation of the time dependent reduced density matrix and the dynamical quantities derived therefrom we refer the reader to appendix B. 

Finally, it is a simple but remarkable difference between our one bath system with two central spins and the homogeneous Gaudin model of a single central spin \cite{SKhaLoss03,BorSt07}, that even if we choose
$\ket{\alpha_2}$ as an $\vec{I}^2$ eigenstate and hence fix $k$ in (\ref{8}) to a single value, due to the higher number of eigenvalues the resulting dynamics can not be described by a single frequency. 

\section{Basic dynamical properties}
\label{dynamics}

We now give an overview over basic dynamical features of the system
considered. Due to the homogeneous couplings, the dynamics of the two central spins can be read off from each other. 
\begin{figure}[h!]
\begin{flushright}
\resizebox{\linewidth}{!}{
\includegraphics{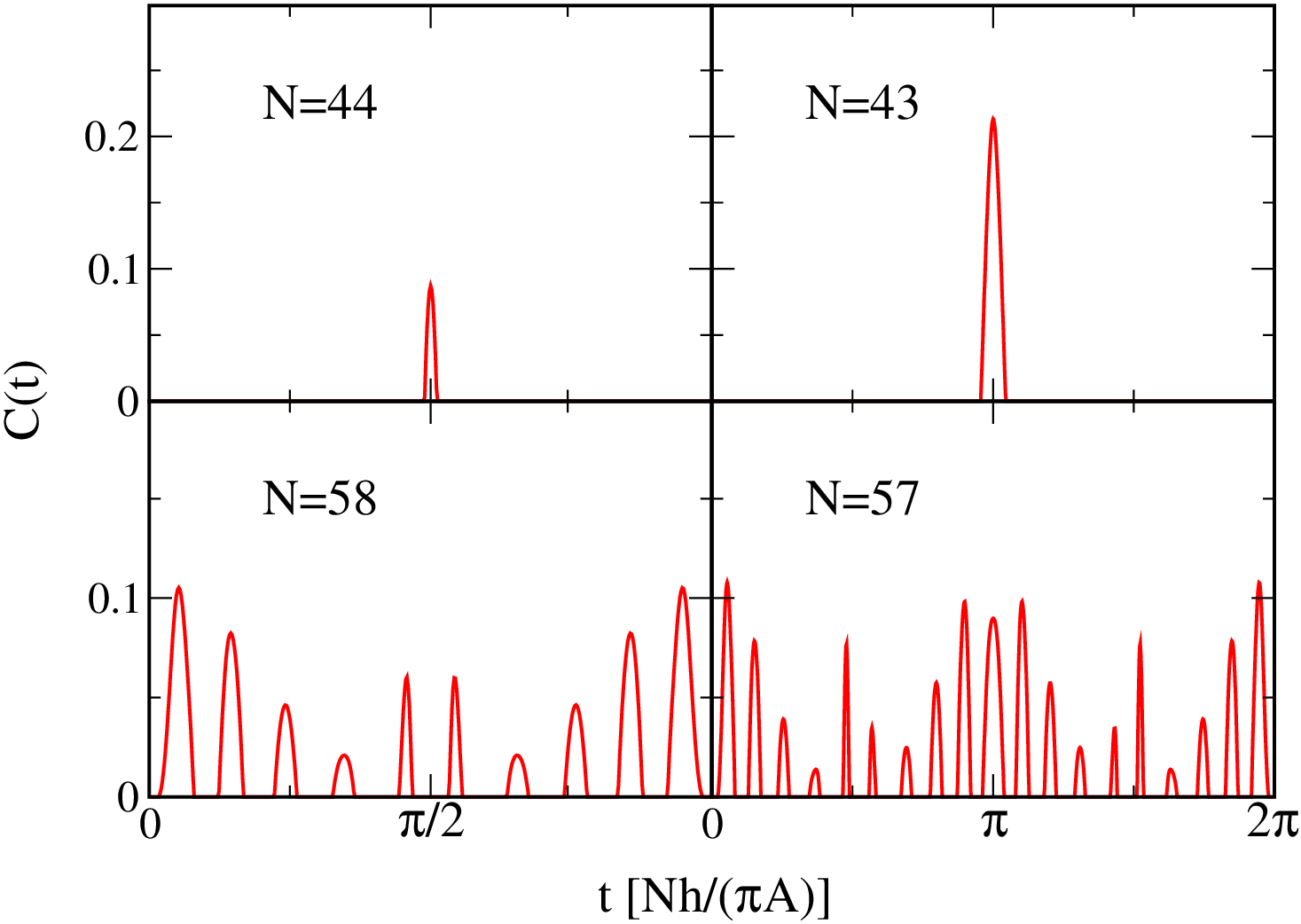}}
\end{flushright}
\caption{\label{Fig:con_hom} (Color online) Entanglement dynamics for $\ket{\alpha_1}=\ket{T_+}$ and $N_D=20$, resulting in $p_b \approx 6 \% - 30 \%$. Instead of an oscillating function we see discrete peaks. Variation of the exchange coupling has no influence because $\ket{T_+}$ is an eigenstate of the central spin coupling term.}
\end{figure}
Hence the following discussion of the dynamics will be restricted to $\langle S^z_1(t) \rangle$. 
\subsection{Electron spin dynamics}

In Figs. \ref{Fig:evenodd1}, \ref{Fig:evenodd2} we consider the completely homogeneous case $\Jx=A/N$ and plot the dynamics for $\ket{\alpha}=\ket{\Uparrow \Downarrow}, \ket{T_+},\ket{T_0}$ and varying polarization $p_b\approx 2\% - 30\%$. A polarization of $30 \%$ does not seem to be particularly high, but the behavior typical for high polarizations
occurs indeed already at such a value.
We omit the singlet case because it is an eigenstate of the system. In Fig. \ref{Fig:evenodd1} the number of spins is even, whereas in Fig. \ref{Fig:evenodd2} an odd number is chosen. Note that we measure the time $t$ in rescaled units $\hbar/(A/2N)$ depending on the 
number of bath spins  \cite{Note1}. Similarly to the homogeneous Gaudin system \cite{SKhaLoss03,BorSt07}, from Figs. \ref{Fig:evenodd1}, \ref{Fig:evenodd2} we see that the dynamics for an even number of spins is periodic with a periodicity of $\pi$ (in  rescaled time units), whereas an odd number of spins leads to a periodicity of $2 \pi$. This is the case for $\Jx$ being any integer multiple of $A/N$. These characteristics can of course be explained by analyzing the level spacings in the different situations. For example, for an even number of bath spins, all level spacings are even multiples of $A/2N$ \cite{Note1}, resulting in dynamics periodic with $\pi$. However, if the number of spins is odd, we get even and odd level spacings (in units of $A/2N$), giving a period of $2 \pi$. For the given case of completely homogeneous couplings the dynamics can be nicely characterized: The number of local extrema for an even number of bath spins within a complete  period, as well as for an odd number of bath spins within half a period, is in both cases given by $N-2N_D+1$. 
\begin{figure}[h!]
\begin{flushright}
\resizebox{\linewidth}{!}{
\includegraphics{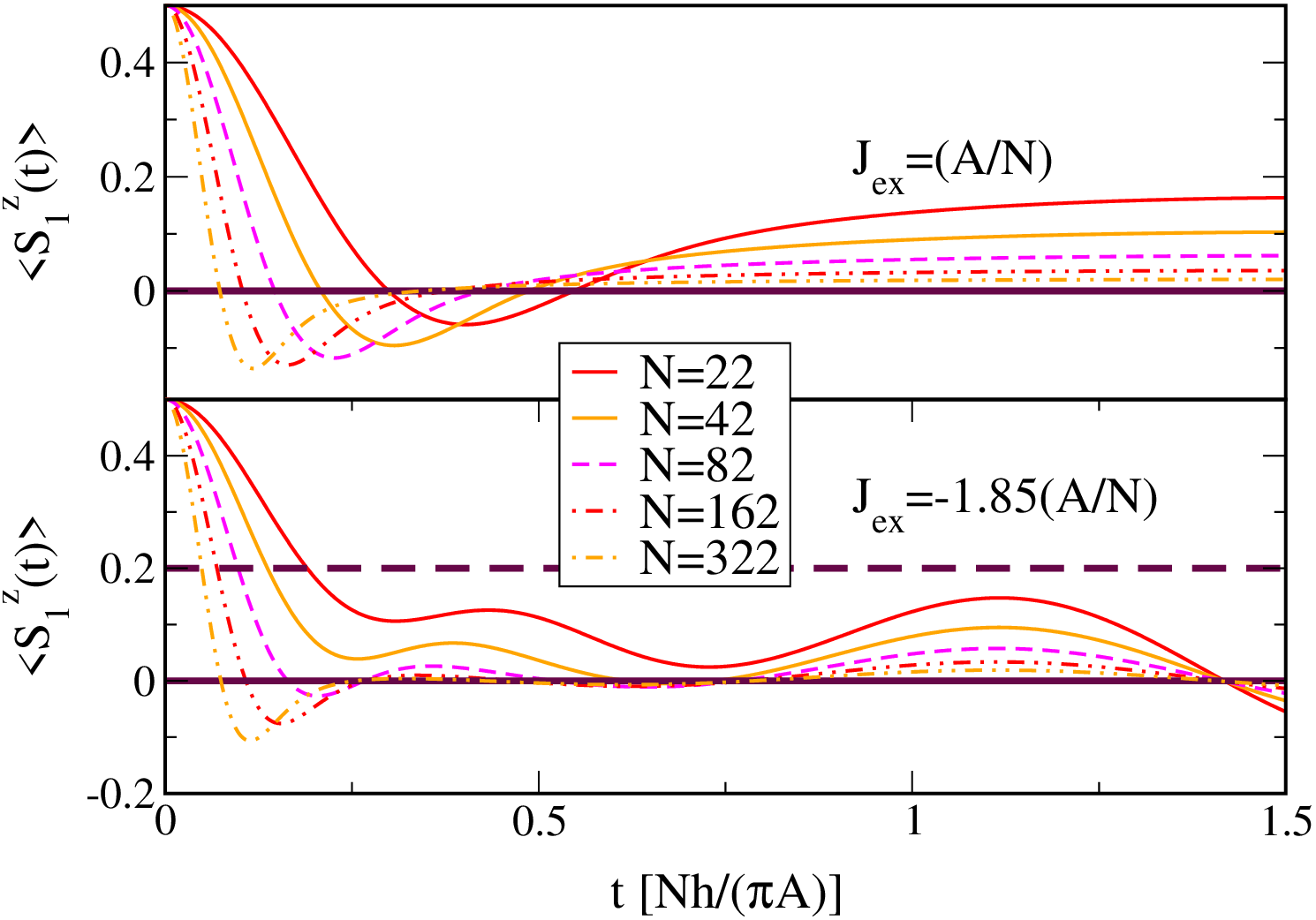}}
\end{flushright}
\caption{\label{Fig:Jsubzero} (Color online) Spin dynamics on short time scales for $\Jx \lessgtr 0$, $p_b=2/N$, and $\ket{\alpha_1}=\ket{\Uparrow \Downarrow}$. The thick solid lines mark the zero level $\langle S^z_1 \rangle=0$ while the thick dashed line (lower panel) represents the
threshold level $\langle S^z_1 \rangle=0.2$ as appropriate
for $\Jx<0$ and small spin baths.                                                                                           }
\end{figure}
This -- so far empirical -- rule holds for all initial central spin states and is illustrated in Figs. \ref{Fig:evenodd1} and \ref{Fig:evenodd2}.

Let us now investigate the spin dynamics for varying exchange coupling,
i.e. the case $\Jx\neq A/N$.
Note that for the initial central spin state
$\ket{\alpha_1}=\ket{T_0}$ this inhomogeneity has no influence on
the spin dynamics since $\ket{T_0}$ is an eigenstate of $\vec{S}_1 \cdot \vec{S}_2$ and 
\begin{equation}
 \left[ H_{\operatorname{hom}},\vec{S}_1 \cdot \vec{S}_2 \right]=0.
\end{equation}
In Fig. \ref{Fig:evenoddJ1} the dynamics for $\ket{\alpha_1}=\ket{\Uparrow \Downarrow}$ and varying exchange coupling is plotted. In the upper two panels we consider the case of low polarization $p_b \approx 10\%$ for an even and an odd number of spins. The remaining two panels show the dynamics for high polarization $p_b \approx 30 \%$. In Fig. \ref{Fig:evenoddJ2} the plots are ordered likewise for a more general linear combination of 
$\ket{\Uparrow \Downarrow}$ and $\ket{T_0}$ ,
$\ket{\alpha_1}=(1/\sqrt{13})\left( 2 \ket{\Uparrow \Downarrow} + 3 \ket{\Downarrow \Uparrow} \right) $. 

From Figs. \ref{Fig:evenoddJ1}, \ref{Fig:evenoddJ2} we see that if the exchange coupling is an odd multiple of $A/2N$, the even-odd effect described above does not occur and we have periodicity of $2 \pi$. In both of the aforementioned situations the time evolutions are symmetric with respect to the middle of the period, which is a consequence of the invariance of the underlying Hamiltonian under time reversal. For a more general exchange coupling, the periodicity, along with the mirror symmetry, of the dynamics is broken on the above time scales.

Considering the case of low polarization, neither the dynamics of initial states with a product nor the one of states with an entangled central spin state dramatically changes if $\Jx$ is varied. However, if the polarization is high, the spin is oscillating with mainly one frequency proportional to $\Jx$. 
\begin{figure}[h!]
\begin{flushright}
\resizebox{\linewidth}{!}{
\includegraphics{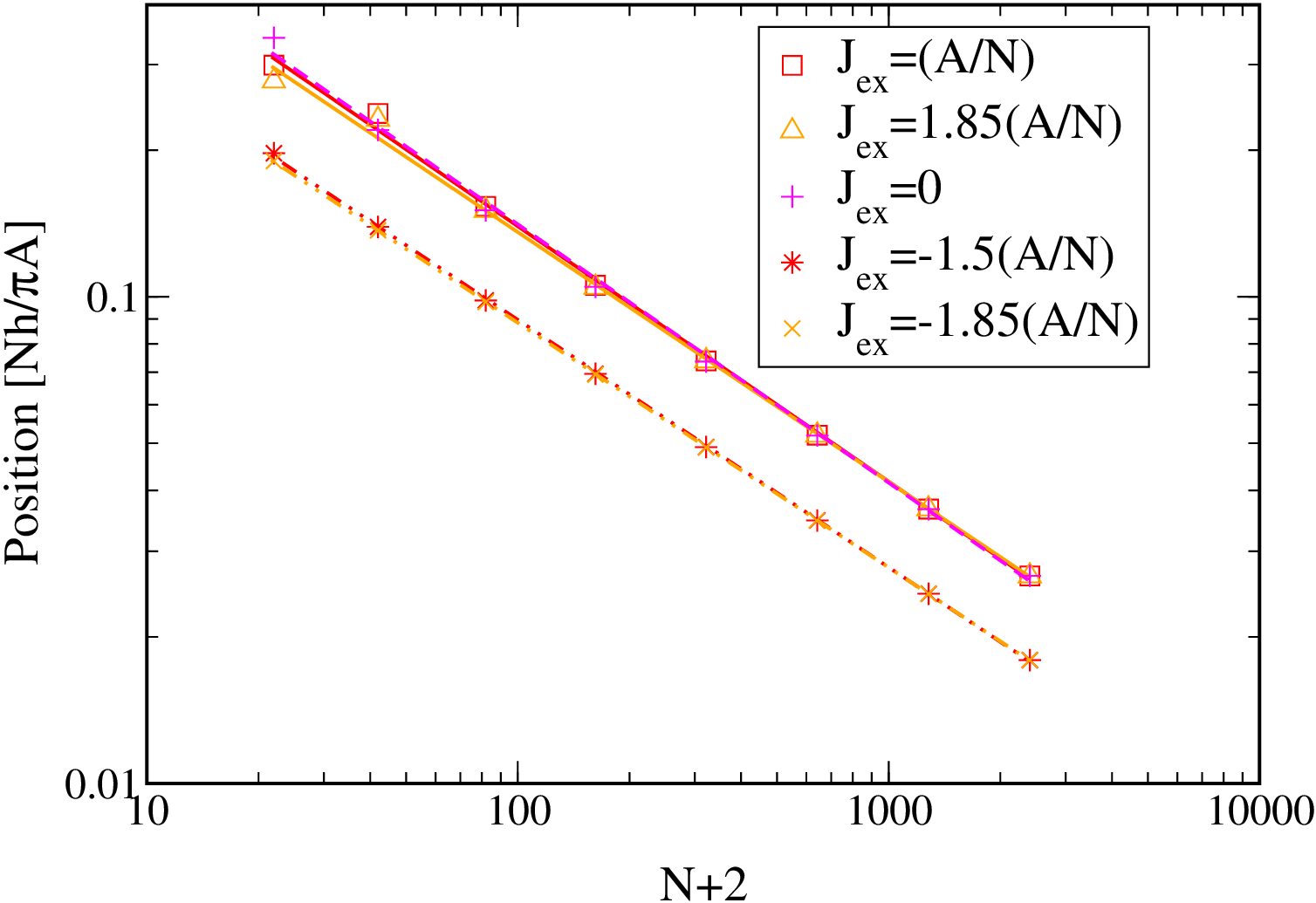}}
\end{flushright}
\caption{\label{Fig:scale} (Color online) Position of the first zero of $\langle S^z_1(t) \rangle$ for $\Jx \geq 0$, and the first intersection with the threshold level $\langle S^z_1 \rangle=0.2$ for $\Jx < 0$, on a double logarithmic scale. We choose $\ket{\alpha_1}=\ket{\Uparrow \Downarrow}$ and a polarization of $p_b=2/N \Leftrightarrow N=2N_D+2$. The curves are fitted to a power law $\propto N^\nu$ with $\nu=-0.52$
($\Jx=(A/N)$), $\nu=-0.51$ ($\Jx=1.85(A/N)$), $\nu=-0.53$ ($\Jx=0$),
$\nu=-0.51$ ($\Jx=-1.5(A/N)$), $\nu=-0.50$ ($\Jx=-1.85(A/N)$).
Note that the parallel offset between the plots for $\Jx \geq 0$ and $\Jx <0$ results from the fact that the intersection with the higher threshold level happens closer to zero.}
\end{figure}
Furthermore the amplitude of the oscillation is larger for the case $\Jx \neq A/N$ than for the completely homogeneous case. This behaviour can be understood as follows: If the polarization is high $d_{N_D} \approx 1$, whereas $d_k \approx 0$ for $k \neq N_D$. 
This means that calculating the spin and entanglement dynamics, we only have to consider the term $k=N_D$. An evaluation of the coeffcients for the different frequencies now shows that the main contribution results from $E_{T_0}-E_S = (A/N)-\Jx$ in obvious notation. Hence if the polarization is more and more increased, this is the only frequency left. If $\Jx=(A/N)$, the two associated eigenstates are degenerate so that in this case the main contribution to the dynamics is constant. This explains why the amplitude of the high polarization dynamics in Figs. \ref{Fig:evenoddJ1}, \ref{Fig:evenoddJ2} is big compared to the one in Figs. \ref{Fig:evenodd1}, \ref{Fig:evenodd2}. For further details the reader is referred to appendix B.

\subsection{Entanglement dynamics}
In Figs. \ref{Fig:con_hom1}, \ref{Fig:con_hom} the concurrence dynamics $C(t)$ for $\ket{\alpha_1}=\ket{\Uparrow \Downarrow}, \ket{T_+}$ is plotted for the same polarizations as in Figs. \ref{Fig:evenoddJ1}, \ref{Fig:evenoddJ2} and varying exchange coupling.

It is interesting that in the second case the concurrence drops to zero for certain periods of time.
This is very similar for the case $\ket{\alpha_1}=\ket{T_0}$ not shown above. As already explained concerning the spin dynamics, the exchange coupling $\Jx$ of course has no influence because $\ket{T_+}$ is an eigenstate of $\vec{S}_1 \cdot \vec{S}_2$. 

It is an interesting fact now that 
for $\ket{\alpha_1}=\ket{\Uparrow \Downarrow}$ and a small polarization changing from $\vert \Jx  \vert > 1$ to $\vert \Jx \vert <1$ increases the maximum value of the function $C(t)$. Furthermore we see from Fig. \ref{Fig:con_hom1} that surprisingly the entanglement is much smaller for the completely homogeneous case $\Jx = A/N$ than for $\Jx \neq A/N$ even for low polarization.

\section{Decoherence and its quantification}
\label{decoherence}

Depending on the choice of the exchange coupling, the dynamics of the one bath model can either be symmetric and periodic or without any regularities. It is now not entirely obvious to determine in how far these dynamics constitute a process of decoherence. Considering for example the spin dynamics for an integer $\Jx$ and an even number of bath spins shown in Fig. \ref{Fig:evenodd1}, one can either regard the decay of the spin as decoherence or, especially due to the symmetry of the function, as part of a simple periodic motion. In Ref.~\cite{BorSt07} the first zero of $\langle S^z_1(t) \rangle$ has been considered as a measure for the decoherence time. In Fig. \ref{Fig:Jsubzero} we illustrate examples of the spin dynamics on short time scales for $\Jx \geq 0$, $\Jx <0$ and a varying number of bath spins. For $\Jx \geq 0$ this procedure is straightforward meaning that $\langle S^z_1(t) \rangle$ crosses the horizontal line $\langle S^z_1 \rangle=0$ before reaching its first
minimum with $\langle S^z_1(t) \rangle<0$.   
However, for $\Jx<0$ and a sufficiently small number of bath spins, as seen from the lower panel of Fig. \ref{Fig:Jsubzero}, such a first minimum is attained before the first actual zero $\langle S^z_1(t) \rangle=0$. This first zero occurs indeed at much large times $t$ whose scaling
behavior as a function of system size $N$ is clearly different from the 
zero positions found for $\Jx \geq 0$, as we have checked in a detailed analysis.
Thus, our evaluation scheme needs to be modified for $\Jx<0$. An obvious way out of
this problem is to either consider large enough spin baths where such an effect does not occur, or to evaluate the intersection with alternative
``threshold level'' $\langle S^z_1 \rangle>0$. In  Fig. \ref{Fig:Jsubzero} we have chosen $\langle S^z_1 \rangle=0.2$, which will be the basis of our following investigation.
As a further alternative, one could also consider the position of the first minimum of $\langle S^z_1(t) \rangle$.
Hence, strictly speaking, it is not per se the first zero of $\langle S^z_1(t) \rangle<0$ which is a measure for the decoherence time, but the scaling behavior of the dynamics on short time scales. 
Following the route described above, in Fig. \ref{Fig:scale} we plot the positions (measured in units of $\hbar/(A/2N)$)
of the first zeroes of $\langle S^z_1(t) \rangle$ for $\Jx \geq 0$, and of the first intersections with the threshold level shown in Fig. \ref{Fig:Jsubzero} for $\Jx<0$, on a double logarithmic scale. We choose a weakly polarized bath $N=2N_D+2\Rightarrow p_b=2/N$, approaching
the completely unpolarized case for $N\to\infty$.
The absolute values of the positions for $\Jx \geq 0$ and $\Jx<0$ differ slightly from each other, which results from the fact that the intersection with the threshold level at $0.2$ happens closer to zero than with the usual threshold level $\langle S^z_1 \rangle=0$. Nevertheless, the scaling behavior is very similar in all cases, and each curve can nicely be fitted by a power law $\propto (N+2)^\nu$ with $\nu\approx -0.5$, a result similar to the one found for the homogeneous Gaudin system 
with only one central spin \cite{BorSt07}.

In a GaAs quantum dot the electron spins usually interact with approximately $N=10^6$ nuclei. Assuming the hyperfine coupling strength to be of the order of $A=10^{-5}$eV, as realistic for GaAs quantum dots \cite{SKhaLoss03}, this results in a time scale of $Nh/(\pi A)= 1.31 \cdot 10^{-4} $s. If we now use the above scaling behaviour $1/\sqrt{N+2}$, we get a decoherence time of $131$ns, which fits quite well with the experimental data \cite{expAwschalom,Koppens05,Petta05,Koppens08}. This is an interesting result not only with respect to the validity of our model: As explained following equation (\ref{flipflop}), generally decoherence results ``directly'' from the electron-nuclear flip-flop terms and through the superposition of product states from the z terms. Above we calculate the decoherence time for $\ket{\alpha_1} =\ket{\Uparrow \Downarrow}$, where the influence of the z terms is eliminated. The fact that we are able to reproduce the decoherence times suggests that the decoherence time caused by the flip-flop terms is equal or smaller than the one resulting from the z parts of the Hamiltonian. It should be stressed that we calculate the decoherence time of an individual electron $T_2$ here. In Ref. \cite{Merkulov02} the decoherence time of an ensemble of dots $T_2^*$ has been calculated  yielding $1$ns for a GaAs quantum dot with $10^5$ nuclear spins.

It is now a well-known fact for the Gaudin system that the decaying part of the dynamics decreases with increasing polarization \cite{SKhaLoss03}. A numerical evaluation shows that this is also the case for two central spins. As explained in the context of Figs. \ref{Fig:evenodd1}, \ref{Fig:evenodd2}, \ref{Fig:evenoddJ1}, \ref{Fig:evenoddJ2} the oscillations of our one bath model become more and more coherent with increasing polarization. Together with the above results for the decoherence this means that, although the homogeneous couplings are a strong simplification of the physical reality, our homogeneous coupling model shows rather realistic dynamical characteristics on the relevent time scales. This is plausible because artifacts of the homogeneous couplings, like the periodic revivals, set in on longer time scales.

\section{Conclusion}

In conclusion we have studied in detail the hyperfine induced spin and entanglement dynamics of a model with homogeneous hyperfine coupling constants and varying exchange coupling, based on an exact analytical calculation.

We found the dynamics to be periodic and symmetric for $\Jx$ being an integer multiple of $A/N$ or an odd multiple of $A/2N$, where the period depents on the number of bath spins. We explained this periodicity by analyzing the level spectrum. For $\Jx=A/N$ we found an empirical rule which charaterizes the dynamics for varying polarization. We have seen that for low polarizations the exchange coupling has no significant influence, whereas in the high polarization case the dynamics mainly consists of one single frequency proportional to $\Jx$. It is not possible to entangle the central spins completely in the setup considered in this
article. 

Following Ref. \cite{BorSt07} 
we extracted the decoherence time by analyzing the scaling behaviour of the first zero. In the case of negative exchange coupling the dynamics strongly changes on short time scales and instead of the first zero we considered the intersection of the dynamics with another threshold level parallel to the time axis. Both cases yield the same result which is in good agreement with experimental data. Hence the scaling behaviour of the short time dynamics can be regarded as a good indicator for the decoherence time. \newline

\acknowledgments
This work was supported by DFG program SFB631. J.~S. acknowledges the
hospitality of the Kavli Institute for Theoretical Physics at the
University of California at Santa Barbara, where this work was reaching completion and was therefore supported in part by the National Science Foundation under Grant No. PHY05-51164.

\appendix 
\section{Diagonalization of the homogeneous coupling model}
The eigenstates of $H_{\operatorname{hom}}$ can be found directly by iterating the well known expressions\cite{Schwabl} for coupling an arbitrary spin to a spin $S=1/2$. Two of these states lie in the triplet sector:
\begin{subequations}
\begin{gather} \label{eig1}
\nonumber \ket{I+1,m,1} = \sqrt{\frac{I+m+1}{2I+2}\cdot \frac{I+m}{2I+1}}\ket{I,m-1}\ket{T_+} \\
\nonumber + \sqrt{\frac{I+m+1}{I+1}\cdot \frac{I-m+1}{2I+1}}\ket{I,m}\ket{T_0}\\
 + \sqrt{\frac{I-m+1}{2I+2}\cdot \frac{I-m}{2I+1}}\ket{I,m+1}\ket{T_-}\\
\nonumber \ket{I-1,m,1}= \sqrt{\frac{I-m}{2I}\cdot \frac{I-m+1}{2I+1}}\ket{I,m-1}\ket{T_+} \\
\nonumber - \sqrt{\frac{I-m}{I}\cdot \frac{I+m}{2I+1}}\ket{I,m}\ket{T_0}\\
 + \sqrt{\frac{I+m}{2I}\cdot \frac{I+m+1}{2I+1}}\ket{I,m+1}\ket{T_-}
\end{gather}
\end{subequations}
As already mentioned in the text, the states are labelled by the quantum numbers $J,m,S$ corresponding to the operators $\vec{J}^2,J^z,\vec{S}^2$. The rest of the quantum numbers due to a certain Clebsch-Gordan decomposition of the bath is omitted. For the eigenstates of the central spin term $\vec{S}_1 \cdot \vec{S}_2$ we used the standard notation:
\begin{subequations}\label{Trip}
\begin{eqnarray} 
 \ket{T_+}&=&\ket{\Uparrow \Uparrow} \\
 \ket{T_0}&=&\frac{1}{\sqrt{2}}\left(\ket{\Uparrow \Downarrow}+\ket{\Downarrow \Uparrow} \right) \\
 \ket{T_-}&=&\ket{\Downarrow \Downarrow} \\
\ket{S}&=&\frac{1}{\sqrt{2}}\left(\ket{\Uparrow \Downarrow}-\ket{\Downarrow \Uparrow} \right) 
\end{eqnarray}
\end{subequations}
The remaining two eigenstates are superpositions of singlet and triplet states. As the expressions are rather cumbersome, it is convenient to introduce the following notation in order to abbreviate the Clebsch-Gordan coefficients:
\begin{widetext}
\begin{eqnarray*}
\left\lbrace  \mu^1_1,\mu^1_2,\mu^1_3,\mu^1_4 \right\rbrace  &=& \left\lbrace  \sqrt{\frac{I+m}{2I}\cdot\frac{I-m+1}{2I+1}},\sqrt{\frac{I+m}{2I}\cdot\frac{I+m}{2I+1}},\sqrt{\frac{I-m}{2I}\cdot \frac{I-m}{2I+1}},\sqrt{\frac{I-m}{2I}\cdot\frac{I+m+1}{2I+1}} \right\rbrace  \\
\left\lbrace  \mu^2_1,\mu^2_2,\mu^2_3,\mu^2_4 \right\rbrace  &=& \left\lbrace  \sqrt{\frac{I-m+1}{2I+2}\cdot \frac{I+m}{2I+1}},\sqrt{\frac{I-m+1}{2I+2}\cdot\frac{I-m+1}{2I+1}},\sqrt{\frac{I+m+1}{2I+2}\cdot \frac{I+m+1}{2I+1}},\sqrt{\frac{I+m+1}{2I+2}\cdot\frac{I-m}{2I+1}} \right\rbrace
\end{eqnarray*}
\end{widetext}
With this definitions the superposition states can be written as:\newline \newline
\begin{eqnarray*}
\ket{1}&=& \mu^1_1 \ket{I,m-1}\ket{T_+}+\frac{\mu^1_3-\mu^1_2}{\sqrt{2}}\ket{I,m}\ket{T_0}\\
&-&\mu^1_4\ket{I,m+1}\ket{T_-}+\frac{\mu^1_3+\mu^1_2}{\sqrt{2}}\ket{I,m}\ket{S}
\end{eqnarray*}
\begin{eqnarray*}
\ket{2}&=& \mu^2_1 \ket{I,m-1}\ket{T_+}+\frac{\mu^2_2-\mu^2_3}{\sqrt{2}}\ket{I,m}\ket{T_0}\\
&-&\mu^2_4\ket{I,m+1}\ket{T_-}-\frac{\mu^2_3+\mu^2_2}{\sqrt{2}}\ket{I,m}\ket{S}
\end{eqnarray*}
These states are degenerate with respect to $H_{\operatorname{hom}}$, hence we are left with the simple task to find a superposition of $\ket{1}$ and $\ket{2}$, which eliminates $\ket{I,m}\ket{S}$. Obviously this is given by
\begin{eqnarray*}
\ket{I,m,1} = \frac{1}{N_T} \left( \frac{\sqrt{2}}{\mu^1_2+\mu^1_3}\ket{1}+ \frac{\sqrt{2}}{\mu^2_2+\mu^2_3}\ket{2}\right),
\end{eqnarray*}
where $N_T= \sqrt{-(I+1)^{-1}+I^{-1}+4}$ is the normalization constant. Inserting $\ket{1}$ and $\ket{2}$ this reads:
\begin{eqnarray}\label{eig2}
\nonumber \ket{I,m,1} &=&  \frac{1}{N_T} \sum_{i=1}^2 \left( \frac{\sqrt{2}\mu^i_1}{\mu^i_2+\mu^i_3}\ket{I,m-1}\ket{T_+} \right. \\
\nonumber &+& (-1)^{i+1}\frac{\mu^i_3-\mu^i_2}{\mu^i_2+\mu^i_3}\ket{I,m}\ket{T_0}\\
&-& \left. \frac{\sqrt{2} \mu^i_4}{\mu^i_2+\mu^i_3}\ket{I,m+1}\ket{T_-} \right) 
\end{eqnarray}
Together with the singlet state 
\begin{equation}\label{eig3}
\ket{I,m,0}=\ket{I,m}\ket{S}
\end{equation}
this solves our problem of diagonalizing the one bath homogeneous coupling Hamiltonian. Furthermore (\ref{eig1}) and (\ref{eig2}) give a solution to the very general problem of coupling an arbitrary spin to a spin $S=1$.

\section{Calculation of the time-dependent reduced density matrix}

Let $H$ be a time-independent Hamiltonian acting on a product Hilbert space $\Hh=\otimes_{i=1}^N \Hh_i$. We denote its eigenvectors by $\ket{\psi_i}$ and the corresponding eigenvalues by $E_i$. In the following we calculate the time-dependent reduced density matrix for an initial state which is a pure state and derive the time evolution $\left\langle O_i(t)\right\rangle$ associated with an operator $O_i$ acting on $\Hh_i$. Then we consider the Hamiltonian (\ref{5}) and give some more details on the corresponding calculations for our model. 

As the eigenstates of $H$ span the whole Hilbertspace $\Hh$, the initial state $\ket{\alpha}$ of the system described by $H$ can be written as 
\begin{equation}
\label{eigwr}
 \ket{\alpha}=\sum_{i} \alpha_i \ket{\psi_i}.
\end{equation}
The time evolution of the initial state results from the application of the time evolution operator $U=e^{-\frac{i}{\hbar}Ht}$. It follows:
\begin{eqnarray}
\label{time}
\nonumber \ket{\alpha(t)}\bra{\alpha(t)}&=&\ket{U \alpha} \bra{U \alpha} \\
\nonumber &=& \sum_{ij} \alpha_i \alpha^*_j \ket{U \psi_i} \bra{U \psi_j} \\
&=& \sum_{ij} \alpha_i \alpha^*_j  e^{-\frac{i}{\hbar} \left(E_i-E_j \right)t } \ket{\psi_i}\bra{\psi_j}
\end{eqnarray}
As $O_i$ acts on $\Hh_i$, the other degrees of freedom have to be traced out
\begin{equation*}
 \rho_i(t)=\Tr_{\Hh \setminus \Hh_i}\left(\ket{\alpha(t)}\bra{\alpha(t)} \right),
\end{equation*}
finally giving the time evolution of the operator:
\begin{equation}
\label{ttime}
 \left\langle O_i(t) \right\rangle = \Tr_{\Hh_i}\left( \rho_i(t) O_i \right) 
\end{equation}
Usually such calculations are done numerically, but for our homogeneous coupling model it is possible to derive exact analytical expressions for the dynamics of the two central spins. 

Following the general scheme, we have to write the initial state in terms of energy eigenstates first. As explained in the text, we consider $\ket{\alpha}=\ket{\alpha_1}\ket{\alpha_2}$, where $\ket{\alpha_1}$ is an arbitrary central spin state and $\ket{\alpha_2}$ is a product state in the bath Hilbertspace $\Hh_N$. Using (\ref{8}) it follows:
\begin{small}
\begin{equation} 
\label{ap21}
\ket{\alpha_1}\ket{\alpha_2}= \sum_{k=0}^{N_D} \sum_{\left\{S_i\right\}} c_k^{\left\{S_i\right\}} \ket{\alpha_1}\ket{\frac{N}{2}-k,\frac{N}{2}-N_D,\left\{S_i\right\}}
\end{equation}
\end{small}
The eigenstates (\ref{eig1}, \ref{eig2}, \ref{eig3}) are given in terms of product states between a basis element from (\ref{Trip}) and an $\vec{I}^2$ eigenstate. Hence we can find the coefficients of (\ref{eigwr}) by solving (\ref{eig1}, \ref{eig2}, \ref{eig3}) for these states and inserting them into (\ref{ap21}). If we arrange the coefficients from (\ref{eig1}, \ref{eig2}, \ref{eig3}) into a $4 \times 4$ matrix $V$ according to
\begin{equation}
\label{V}
V=\left( \begin{array}{c|cccc}
&\ket{T'_+}&\ket{T'_0}&\ket{T'_-}&\ket{S'}\\\hline
\ket{I+1,m,1}&\ddots&&& \\
\ket{I,m,1}&&\ddots&&\\
\ket{I-1,m,1}&&&\ddots&\\
\ket{I,m,0}&&&&\ddots
\end{array}\right),
\end{equation} 
this is simply done by transposing $V$. Here $\ket{T'_+}=\ket{I-1,m}\ket{T_+}$ and analogously for the other states. In order to abbreviate the following expressions we denote the energy eigenstates by $\ket{\psi_i}$ as in the general considerations above and number with respect to (\ref{V}). Analogously we introduce the shorthand notation $\ket{i}$ for the basis states (\ref{Trip}). 

In order to avoid further coefficients we choose $\ket{\alpha_1}$ to be the $j$-th element of (\ref{Trip}) and find the following expression for the decomposition of the initial state into energy eigenstates
\begin{equation}
\label{dec}
 \ket{j}\ket{\alpha_2}= \sum_{l=1}^4 \sum_{k=0}^{N_D} \sum_{\left\{S_i\right\}} c_k^{\left\{S_i\right\}} V^T_{jl} \ket{\psi_l},
\end{equation}
where it is has to be noted that the elements $V^T_{jl}$ and the eigenstates $\ket{\psi_l}$ depent on the quantum numbers the sums run over. Hence in our case the coefficients $\alpha_i$ and the eigenstates $\ket{\psi_i}$ in fact have more than one index. \newline
Inserting (\ref{dec}) and (\ref{eig1}, \ref{eig2}, \ref{eig3}) in (\ref{time}) and tracing out the bath degrees of freedom, we finally arrive at the reduced density matrix of the two central spins
\begin{gather}
\label{rho}
 \rho(t):= \Tr_{\Hh_N}\left( \ket{\alpha(t)}\bra{\alpha(t)} \right)=  \\
\nonumber \sum_{k=0}^{N_D} \underbrace{\sum_{\left\lbrace S_i\right\rbrace }\left( c_k^{\left\lbrace S_i\right\rbrace }\right)^2 }_{d_k} \sum_{l,m,n,o=1}^4 V_{jl}^T V_{jm}^T V_{ln} V_{mo} e^{-\frac{i}{\hbar}\left(E_l -E_m \right)t }\ket{n}\bra{o}.
\end{gather}
If we now choose $O_1 = S^z_1$, we have to trace out the second central spin. Inserting the result into (\ref{ttime}) then gives rise to the time evolution $\left\langle S^z_1(t) \right\rangle $. This is given by (\ref{rho}) with $n=o$, multiplied by coefficients resulting from the eigenvalues of $S^z_1$. As mentioned in the text, for high polarizations $d_k \approx 0$ if $k \neq N_D$. Fixing $l,m$ we can calculate the contribution of the respective frequency by evaluating the remaining sum over $n$. If the polarization is strongly increased, all frequencies are suppressed except for $E_2-E_4=E_{T_0}-E_S$.

\end{document}